# Canonical Quantization Approach of a Class of a Dissipative System: Applications to Quantum Tunnelling with Dissipative Coupling

N. Emir Anuar

Independent Researcher, Putrajaya, Malaysia

**Abstract**

We present a formalism for which a dissipative system is given by a variational principle. The formalism applies to dynamical systems where its trajectory is strictly monotonic. Subsequently, we derive its Lagrangian and Hamiltonian functions. From the Hamiltonian, we quantize canonically the classical particle in a viscous media. We study the free quantum particle in a viscous media and the dissipative quantum tunnelling. It is found that the dissipation influences tunnelling probability by a factor of $\exp(-\eta \Delta q^2/\hbar)$, where $\Delta q$ is the barrier width, close to the result of Caldeira and Leggett.

## 1. Introduction

Dissipative systems seems to be inherent in nearly all classical systems. Philosophically, it is an open question whether nature, in its purest form, is dissipative. Dissipation, after all, may be reflecting on our inability to model nature in its entirety, for the complexity will be unbearable for most but the simplest of systems. Conservative systems can be subjected to a more abstract analyses through the mechanism of variational principles. The beauty of the variational principles, whereby motion follows the principle of least action, is perhaps a principle of nature surpassed by none when elegance is of concern. Dissipative systems, on the other hand, are not given by a variational principle [1,2], and thus may be seen, mathematically, as perturbation around conservative systems. Extensive literature exists discussing quantization of dissipative systems, indicating the potential importance of the subject. Today, the need is real, as quantum mechanical systems are more common-place than ever, coupled with the fact that they are approaching scale where coupling to the environment is becoming more significant [3]. It is a curious fact that most observable dynamical systems are subjective to a certain extent, dissipative coupling. Yet the seemingly fundamental laws do not naturally include dissipation.

Various methods have been employed in quantizing dissipative systems, but there is no single methodology that is accepted as the primary method. Typical formulation of quantum mechanics for a given system starts by defining its Lagrangian or the Hamiltonian. This is a major hurdle for dissipative system, as Bauer showed in his paper that dissipative systems are not given by the variational principles [4]. A novel approach was suggested by Riewe [5,6], in which he used fractional derivative to bypass Bauer's Corollary, giving rise to interesting structures of mechanics. However, such approach increase the computational difficulty at the offset. Alternatively, one could employ the machinery of inverse-problems [7 - 10], where one can actually create formalisms to derive Lagrangian functions from equations of motion. This problem is studied extensively in the context of bypassing Bauer's Corollary, where, similarly to the philosophy taken by Riewe, the Lagrangians do not depend explicitly on velocity. These approaches however will distort the physical interpretation of the Hamiltonian and the canonical momenta, which may be acceptable in some cases. Some authors started directly constructing a quantum mechanical Hamiltonian which in the classical limit gives the mass-spring-damper system. The Caldeira – Leggett Hamiltonian [14] is one example of such construct. It provides exemplary benefits and applicability to actual systems, which is evident by the ubiquity of the formulation in recent times, with its extensive use in condensed matter physics.

In this paper, we will attempt to provide a canonical quantization for a limited set of classical dissipative system. A major restriction of the methodology outlined here is the requirement that motion has a trajectory that is strictly monotonic. If there is no driving force, then the force that is acted upon is the dissipative force, which by definition, slows down the particle. One such example of a classical system that falls within the requirements of this formulation is the over - damped spring system. We shall provide as much formalization of systems that may be admitted by the current formulation.

The formulation developed will then be applied to a quantum tunnelling problem. Although we do not arrive at any testable quantitative prediction, we obtain a rather striking qualitative conclusion. The main appeal of our result, in our humble opinion, is the similarity of the suppressing term to the ones obtained by Caldeira and Leggett. This paper does not maintain a strict mathematical rigor, but care is taken where necessary. Dynamical systems are all one – dimensional.

## 2. Preliminaries

Consider a motion of a particle with an initial velocity that is moving with friction dependent on velocity. Intuitively, we would guess that the system will eventual come to rest. Hence we know that the kinetic energy of the system is non-increasing. Further intuition would also suggest that the trajectory and its velocity will be monotonic functions.

These notions are the fundamental elements in the current construct, and need formalization before proceeding further.

**Strictly Dissipating Systems**

We would like to formalize a type of dynamical system that will be the focal point of this paper. As previously explained, our formulation is applied to dissipative systems that have trajectory that are monotonic.

**Definition 1.0.** Consider a system with energy $E = T + V$, where T is the kinetic energy and V is the balance of energy. We define kinetic energy as follows;

$$T = \frac{1}{2}mv^2 \tag{1}$$

A system is strictly dissipative (SD) if for the interval of time $t_a$ and $t_b$, $t_b > t_a$, then $T_{tb} < T_{ta}$. Correspondingly, the source energy will be strictly increasing for within the same interval.

In our formalism, we shall define dissipative energy in the usual sense as part of the overall energy, accounted for in the balance of energy term. A dissipative "potential function," or rather the dissipative force generator or source function, is the Rayleigh dissipation function that is integrated over the time period of interest.

Dissipative energy are commonly represented by an infinite degree of freedom harmonic oscillators as the so-called heat bath. Here, due to re-parameterization as will be demonstrated later, the energy is still a constant. However, this poses a problem that energy may flow back into the kinetic component of energy.

Several lemmas can directly be derived from the definition above and we shall provide the proofs of these lemmas in the appendix.

**Lemma 1 (Monotonicity of solution).** If a system is SD, the trajectory of the system is strictly monotonic.

**Lemma 2 (Vanishing of Kinetic Energy at sufficiently large time).** if the system is SD, then lim(t -> 0) T = 0.

**Lemma 3 (Non-decreasing source energy).** If the system is SD, then the balance of energy is non-decreasing.

## 3.     Dissipative potential system

Inasmuch there is a lemma proving that dissipative system is not given by a variational principle, there exist a method where dissipative systems can be included in the Lagrangian equation of motion, as Lord Rayleigh showed more than a century ago. The dissipative force is given on the right hand side of the Langrage equation of motion, as follows;

$$\frac{d}{dx}\frac{\partial L}{\partial x} - \frac{\partial L}{\partial x} = -\frac{\partial R}{\partial q} \tag{2}$$

Where R is the Rayleigh dissipation function;

$$R = \frac{1}{2}\sum_a K_{jk} v_a^j v_a^k \tag{3}$$

In one dimension, we can write

$$R = \frac{1}{2}k\dot{q}^2 \tag{4}$$

This serves as the generating, or potential, function for the following force;

$$Q = \frac{1}{2}\sum_a \nabla_v R \frac{\partial v}{\partial \dot{q}^k} = -\frac{\partial R}{\partial \dot{q}^k} \tag{5}$$

from which the following force description, in the case of one dimension, can be derived;

$$Q = -k\dot{q} \tag{6}$$

It is observed that the Rayleigh dissipation function is twice the power lost from motion due to dissipation;

$$-Q_k^{(nc)}\dot{q}^k = P = 2R \tag{7}$$

Therefore, the Rayleigh dissipation function relates to the rate of energy associated by the dissipative coupling through;

$$\frac{dW}{dt} = 2R \tag{8}$$

Thus,

$$\frac{dW}{dt} = k\dot{q}^2 \tag{9}$$

Energy dissipated through time $t_a$ and $t_b$ is therefore given as;

$$W = \int_{t_a}^{t_b} k\dot{q}^2 \, dt \tag{10}$$

$$W = \int_{t_a}^{t_b} \int_{\dot{q}_{t_a}}^{\dot{q}_{t_b}} dt \, d\dot{q} \, Q(\dot{q}, t) \tag{11}$$

The last equation describes energy from dissipated influences, or the so-called the heat bath. The heat – bath is an energy pool where it is prohibited physically for the energy to flow back into the remainder of the energy component of the system. In the case of a free particle in a viscous media, the energy in the heat bath cannot contribute to the increase of kinetic energy. Hence without external energy, the particle will always tend to go to rest.

An equation of motion with a force term given as above, according to Bauer's corollary, is not given by a variational principle. This is our point of departure, where we now explore a potential change in variable. Since the Rayleigh dissipation acts as a "generator" of our force term in the eventual equation of motion, we would like to introduce the term "source function" to denote Rayleigh dissipation function. This is purely on lexical purpose; dissipation is associated with energy moving out of system, but our definition of energy and the Hamiltonian includes the dissipated energy term, or the heat bath.

Suppose we have the solution of the equations of motion in the form,

$$q = \gamma(t) \tag{12}$$

Since trajectory is monotonic, it is both surjective and bijective, ensuring the existence of its inverse function;

$$t = \gamma^{-1}(q) \tag{13}$$

Differentiating once the trajectory function yields the velocity, and if we substitute the inverse (number) into the derivative;

$$\dot{q} = \gamma' \circ \gamma^{-1}(q) \tag{14}$$

Hence we have a re - parameterized velocity as a function of position. If write $\gamma' \circ \gamma^{-1} = h$, and substitute the above re-parameterization into the source function, we will arrive at the following;

$$W = \int_{t_a}^{t_b} \int_{\dot{q}_{t_a}}^{\dot{q}_{t_b}} dt\, d\dot{q}\; Q(h(q), t) \tag{15}$$

Intuitively, we are tempted to write;

$$W \equiv \int_{q_a}^{q_b} dq\; Q(h(q), t) \tag{16}$$

If $Q(h(q), t) = H(q, t)$. Then we have;

$$\int_{t_a}^{t_b} \int_{\dot{q}_{t_a}}^{\dot{q}_{t_b}} dt\, d\dot{q}\; Q(h(q), t) \equiv \int_{q_a}^{q_b} dq\; Q(h(q), t) \tag{17}$$

It is our next objective to show, if the system if strictly dissipative, the identity above is indeed an equivalance.

**Equivalence of transformed source function**

We would require to define the existence of two integrals in order to proof the equivalence:

**Definition One (Existence of power dimensioned integral of force dependent on velocity).**

The integral of $\int F(\dot{q}, t)\, d\dot{q}$ is defined when the following limit exists;

$$\int f(\dot{q})\,d\dot{q} = \lim_{n\to\infty}\sum_{i=0}^{n} F(\xi)[\dot{q}(t_{i+1}) - \dot{q}(t_i)] \tag{18}$$

for $t_i \leq \xi \leq t_{i+1}$.

**Definition Two (Existence of time integral of power, with dimension of energy).**

The integral of $\int P(\dot{q}, t)\,dt$ is defined when the following limit exists;

$$\int P(\dot{q}, t)\,dt = \lim_{n\to\infty}\sum_{j=0}^{n} f(\varsigma)[t_{j+1} - t_j] \tag{19}$$

for $t_j \leq \xi \leq t_{j+1}$.

**Theorem 1.0 (Equivalence of Integral Transformation).** Given that a bijection $q = \gamma(t) \in C'$ exists and $h(q) = \dot{\gamma} \circ \gamma^{-1}(q)$, then

$$\int_{t_a}^{t_b}\int_{\dot{q}_{t_a}}^{\dot{q}_{t_b}} dt\,d\dot{q}\, Q(h(q), t) = \int_{q_a}^{q_b} dq\, Q(h(q), t) \tag{20}$$

The integral exist for $q = \gamma(t) \in C'$, $t_a \leq t \leq t_b$.

Proof. We write,

$$W = \int_{t_a}^{t_b}\int_{\dot{q}_{t_a}}^{\dot{q}_{t_b}} dt\,d\dot{q}\, Q(h(q), t) \tag{21}$$

If we write the integral is a Riemann sum, we shall arrive at,

$$W = \lim_{m\to\infty}\lim_{n\to\infty}\sum_{i=0}^{m}\sum_{j=0}^{n} F(\xi)[\dot{q}(t_{i+1}) - \dot{q}(t_i)][t_{j+1} - t_j] \tag{22}$$

for $\dot{q}(t_j) \leq \xi \leq \dot{q}(t_{j+1})$. We are given a bijection $q = \gamma(t) \in C'$, hence we have $h(q) = \dot{\gamma} \circ \gamma^{-1}(q)$ as stated above. Thus, the range of integration can be also written as $h(t_j) \leq \varsigma \leq h(t_{j+1})$, which corresponds to $q_j \leq \varsigma \leq q_{j+1}$. If we commute the time difference and its associated limit, we will arrive at;

$$W = \lim_{m\to\infty}\lim_{n\to\infty}\sum_{i=0}^{m}\sum_{j=0}^{n} F(\xi)[\dot{q}(t_{i+1}) - \dot{q}(t_i)][t_{j+1} - t_j] \tag{23}$$

If we take the limit of n, the expression above, using **Definition 2**, becomes;

$$W = \lim_{m\to\infty}\sum_{j=0}^{m} f(\varsigma)[q_{j+1} - q_j] \tag{24}$$

If we now take the limit of m, we will finally arrive at the expression;

$$W = \int_{q_a}^{q_b} dq \; Q(h(q), t) \tag{25}$$

Therefore, we have established the equivalence of the two integrals.

∎

**Example as applied to Strictly Dissipative system**

According to the theorem above, a source function transformation via re-parameterization is given as;

$$W(q,t) = \iint d\dot{q} \, dt \; G(\dot{\gamma} \circ \gamma^{-1}(q), t) \tag{26}$$

The following particle in a viscous media system is given by the following equation

$$m\ddot{q} + \eta \dot{q} = 0 \tag{27}$$

Interestingly, re-parameterization only requires integrating the above, once with respect to time, yielding;

$$\dot{q} = -\frac{\eta}{m} q \tag{28}$$

If we apply this equation into the force term, we will have;

$$Q = \frac{\eta^2}{m} q \tag{29}$$

Which will result in the following equation of motion;

$$m\ddot{q} - \frac{\eta^2}{m} q = 0 \tag{30}$$

Accordingly, by virtue of Theorem 1.0, the dissipative source function can be written as;

$$W = \frac{\eta^2 q^2}{2m} \tag{31}$$

From where a Lagrangian can be formulated, hence given by a variational principle.

As a matter of completeness, we shall verify that the transformed equation indeed gives the same trajectory as the original description of the dynamical system. The original system has the equation

$$m \ddot{q}(t) + \eta \, \dot{q}(t) = 0 \tag{32}$$

which admits a solution of the form,

$$q(t) = C_1 \exp(r_1 t) + C_2 \exp(r_2 t) \tag{33}$$

The constants $C_1$ and $C_2$ are given by the boundary conditions. The constants $r_1$ and $r_2$ are given by,

$$r_{1,2} = -\frac{1}{2}\left(\frac{\eta}{m} \pm \sqrt{\frac{\eta^2}{m}}\right) \tag{34}$$

$$r_{1,2} = -\frac{\eta}{m}, 0 \tag{35}$$

Thus, the trajectory of the system q(t) is given by,

$$q(t) = C_1 \exp\left(-\frac{\eta}{m}t\right) \tag{36}$$

On the other hand, using the transform given by Theorem 1.0, the equation of motion is now given by,

$$m\ddot{q} - \frac{\eta^2}{m}q = 0 \tag{37}$$

$$\ddot{q} - \frac{\eta^2}{m^2}q = 0 \tag{38}$$

The transformed system admits a similar solution as the original system, but the constants $r_1$ and $r_2$ are given by,

$$r_{1,2} = \pm\frac{1}{2}\sqrt{\frac{4\eta^2}{m^2}} \tag{39}$$

$$r_{1,2} = \pm\frac{\eta^2}{m^2} \tag{40}$$

If we insist that the system is strictly dissipative, then we must only admit the term with negative sign. Hence the solution is identical to the original system.

## 4. Canonical Quantization utilizing Transformed Source Functions

**Applications to Free Particle in a viscous media**

We have developed a formalism from which a strictly dissipative dynamical system can be derived from a variation. One direct consequence of this is the Lagrangian can be found, and subsequently its Hamiltonian can be calculated. Consider the system:

$$m\ddot{q} - \frac{\eta^2}{m}q = 0 \tag{41}$$

Using Theorem 1.0, the source function is given by;

$$W = \frac{\eta^2 q^2}{m} \tag{42}$$

In the energy representation, the source function is the measure of the energy taken away from the initial energy of the system, in which our case is assumed to be kinetic. In that sense, the source function is fundamentally a heat bath. Apart from re-parameterization requirement, a main reason, intuitively, why this formalism applies only to strictly dissipative system is to avoid energy flowing back into the kinetic form. In that respect, our formulation is necessarily incomplete, but applicable only to strictly dissipative system. Therefore, the Lagrangian for the system (number) is;

$$L = \frac{1}{2} m \dot{q}^2 - \frac{\eta^2 q^2}{2m} \tag{43}$$

A simple check by applying the Lagrangian to its equation shows that the equations of motion is recovered. In the definition of strictly dissipative system, where the Lagrangian above is a member of, the system is closed, whereby the energy is constant. Therefore, the Hamiltonian is equal to the energy;

$$E = \frac{p^2}{2m} + \frac{\eta^2 q^2}{2m} \tag{44}$$

The Schrodinger's Equation is given by;

$$\hat{H}\psi = \hat{E}\psi \tag{45}$$

Therefore, if we promote both the Hamiltonian and energy into operators, such that;

$$p \to \hat{p} = i\hbar \frac{\partial}{\partial x}, E \to \hat{E} = i\hbar \frac{\partial}{\partial t}, q \to \hat{q} = q \tag{46}$$

then it is rather direct to construct the Schrodinger's Equation for the system in question. Since the energy, by definition, is constant, we can utilize the time independent Schrodinger's Equation by assuming the wave-function to be of the form;

$$\psi(x, t) = \varphi(t)\phi(x) \tag{47}$$

By substituting the operators above into the Hamiltonian;

$$-\frac{\hbar}{2m} \frac{\partial^2 \phi(x)}{\partial x^2} + \frac{\eta^2 x^2}{2m} \phi(x) = E\phi(x) \tag{48}$$

where the time evolution is given by;

$$\varphi(t) = \exp(-{iEt}/{\hbar}) \tag{49}$$

This equation should explain the quantum mechanical behaviour of a one-dimensional particle in a viscous media. It is to our advantage that the equation takes the form of the harmonic oscillator, whereby its solution is well studied and understood. We shall apply one of a common ways of solving this type of problem, and derive the solution for equation. We write;

$$\frac{\partial^2 \phi(x)}{\partial x^2} + \left(\frac{2mE}{\hbar^2} - \frac{\eta^2 x^2}{\hbar^2}\right)\phi(x) = 0 \tag{50}$$

This solutions has the general solution of the form;

$$\phi_n(x) = A \exp\left(-\frac{1}{2}\frac{\eta}{\hbar}x^2\right) H_n\left(x\frac{\eta}{\hbar}\right) + B \exp\left(\frac{1}{2}\frac{\eta}{\hbar}x^2\right) H_n\left(x\frac{\eta}{\hbar}i\right) \tag{51}$$

We discard the second term by setting B to zero, since the exponential blows up at infinity, disallowing normalization of the wave - function. The energy eigenvalues are given by, for integer n;

$$E_n = \frac{\eta\hbar}{2m}(2n+1) \tag{52}$$

The full wave-function is given by;

$$\psi(x,t) = N_n \exp\left(-\frac{1}{2}\frac{\eta}{\hbar}x^2\right) H_n\left(x\frac{\eta}{\hbar}\right) \exp(-iE_n t/\hbar) \tag{53}$$

where $N_n$ is the normalization constant.

The wave - function has the characteristics vanishing property as position goes further than the origin. Furthermore, the suppressing term is $\exp\left(-\frac{1}{2}\frac{\eta}{\hbar}x^2\right)$, which has the form of the suppression factor due to damping in the problem of tunnelling as solved by Caldeira and Leggett. Time characteristics is not affected by definition of our system. It is interesting to see the expectation value problems if we separate kinetic and heat bath components of the energy. We hope to apply these formulation to derive the time-evolution of the kinetic energy, and understand the fully the dissipation effect on the wave-function. These statements are, however, not substantiated in the present work.

**Quantum Tunnelling of Particle with Energy Barrier Dissipative Coupling**

A more relevant problem that may have opportunities for physical tests is the tunnelling problem. It is a well-known fact that macroscopic quantum tunnelling does happen in a number of physical circumstances. A first concrete theoretical examination of this phenomenon is perhaps given by Caldeira and Leggett in 1981 [14], where they concluded qualitatively that the dissipation causes the tunnelling amplitude by an exponential factor dependent on the dissipative coupling term.

The quantum tunnelling problem, in our example, will be stated as follows [13]: consider a region along the horizontal axis whereby the potential energy is given as such;

$$V(x) = \begin{cases} 0, & -\infty < x < 0 \\ V_B, & 0 < x < \Delta q \\ 0, & \Delta q < x < +\infty \end{cases} \quad (54)$$

This represents a rectangular barrier of height $V_B$ separating the regions ($-\infty < x < 0$) and ($\Delta q < x < +\infty$). We assume that dissipative coupling happens within the barrier, and the particle is approximated as free particle outside the barrier. The free-particle solution applies to both regions before and after the barrier. Within the barrier, we have the following Schrodinger's equation for the zeroth level energy;

$$\phi_n(x) = C \exp\left(-\frac{1}{2}\frac{\eta}{\hbar}x^2\right) \quad (55)$$

Thus, the wave-function in the horizontal axis as introduced earlier are given by;

$$\phi(x) = \begin{cases} A_I \exp(ikx) + A_R \exp(-ikx), & -\infty < x < 0 \\ B \exp\left(-\frac{1}{2}\frac{\eta}{\hbar}x^2\right), & 0 < x < \Delta q \\ A_T \exp(ikx), & \Delta q < x < +\infty \end{cases} \quad (56)$$

Probability amplitude of quantum tunnelling, T, by definition, is the ratio between the magnitude of the wave after the barrier and the magnitude of the incident wave. Mathematically, we can state that

$$T = \frac{|A_T|^2}{|A_I|^2} \quad (57)$$

We shall attempt to define this equation by expressing both its elements in terms of the constant B. We shall solve the constants $A_I$, $A_R$, $A_T$ and B such that $\phi(x)$ and $\phi'(x)$ are continuous at $x = 0$ and $x = \Delta q$. Continuity at $x = 0$ requires;

$$A_I + A_R = B \text{ and } A_I ik - A_R ik = -\frac{B\eta}{\hbar} \quad (58)$$

Meanwhile, continuity at $x = \Delta q$ necessitates the relations;

$$B \exp\left(-\frac{1}{2}\frac{\eta}{\hbar}\Delta q^2\right) = A_T \exp(ik\Delta q), \text{ and } -\frac{B\eta}{\hbar}\exp\left(-\frac{1}{2}\frac{\eta}{\hbar}\Delta q^2\right) \quad (59)$$
$$= A_T ik \exp(ik\Delta q)$$

With some straight-forward algebraic manipulation, we have;

$$|A_T|^2 = \frac{-\frac{\eta}{\hbar}\exp\left(-\frac{\eta}{\hbar}\Delta q^2\right)}{ik \exp(ik\Delta q)}|B|^2 \quad (60)$$

and

$$|A_I|^2 = \frac{1}{4\hbar k^2}(\eta^2 + \hbar^2 k^2)|B|^2 \quad (61)$$

Therefore, the transition amplitude is given by;

$$T = \frac{\left(\dfrac{-\frac{\eta}{\hbar}\exp\left(-\frac{\eta}{\hbar}\Delta q^2\right)}{ik\exp(ik\Delta q)}|B|^2\right)}{\left(\dfrac{1}{4\hbar k^2}(\eta^2 + \hbar^2 k^2)|B|^2\right)}$$

$$= \frac{4\eta^2 \exp\left(-\frac{\eta}{\hbar}\Delta q^2\right)\exp(2k\Delta q)}{\eta^2 + \hbar^2 k^2}$$

$$= \frac{4}{(1 + \hbar k^2/\eta^2)}\exp\left(-\frac{\eta}{\hbar}\Delta q^2\right)\exp(2k\Delta q)$$

$$= 4\left(1 + \frac{2m\hbar}{\eta}(E - V_B)\right)^{-1}\exp\left(-\frac{\eta}{\hbar}\Delta q^2\right)\exp(2k\Delta q) \qquad (62)$$

Although we won't attempt here to obtain exact numerical value, the qualitative result is clear; the probability amplitude is suppressed by the dissipation factor through the term $\exp(-\eta/\hbar \; \llbracket \Delta q \rrbracket ^2)$. This mirrors the result of Caldeira and Leggett. Interestingly, they calculated the amplitude from a quantum mechanical Hamiltonian function which was shown to give raise to the classical damped harmonic oscillator in its classical limit, whereby our calculations derived from a classical analogue.

Another point of appeal is the fact that both the Hamiltonian and Lagrangian approach give rise to similar qualitative result, for Caldeira and Leggett used the path integral (Lagrangian based) method and in this paper, we quantize canonically using a classical Hamiltonian of the classical system.

## 5. Conclusions

A canonical quantization is presented for a particle moving in viscous media, and we have found that it gives similar qualitative result as with path integral quantization, in that both give rise to a similar exponential factor dependent on the dissipative parameter. Fundamentally there is nothing new presented here, except for a point of view into an old, outstanding problem.

It should be stressed again here that this formulation is incomplete. Its main objective is to show qualitative similarity of treating dissipative system in a slightly different manner. Furthermore, the current formulation is valid within a certain very strict conditions, eliminating its application to most but only a certain few system. That said, it is hoped that the narrow scope of applicability will eventually yield beneficial results.

Finally, a final test of validity is experimental tests. We hope to provide one in a later work. Direct applications in condensed matter phenomenon is abundant, and we hope to make use of the current formulation in dissipation problems in superfluidity.

## A1. Appendix

Here we give proofs to the lemmas introduced in section 2.

**Lemma 1 (Monotonicity of solution):** If a system is SD, the trajectory of the system is strictly monotonic.

Since by definition T is strictly non-increasing,

$$v_2 < v_1 \tag{A.1}$$

Hence

$$\frac{v_2 - v_1}{\Delta t} < 0 \tag{A.2}$$

Let $v_2 - v_1 = \Delta v$, then we have;

$$\lim_{\Delta t \to 0} \frac{v_2 - v_1}{\Delta t} < 0 \tag{A.3}$$

$$\dot{v} < 0 \tag{A.4}$$

If take the derivative of energy, and equate it to 0,

$$\frac{\partial E}{\partial v}\dot{v} + \dot{V}_D = 0 \tag{A.5}$$

$$mv\dot{v} + \dot{V}_D = 0 \tag{A.6}$$

By definition of SD systems,

$$mv\dot{v} < 0 \tag{A.7}$$

Since $\dot{v} < 0$, v must be always be positive throughout the motion. Therefore no reversal in direction of motion exist throughout the interval. Thus $q(t)$ is strictly monotonic.

**Lemma 2 (Vanishing of Kinetic Energy at sufficiently large time):** if the system is SD, then lim(t -> 0) T = 0.

Proof. If we write

$$\int_0^\infty \dot{T}\, dt = A \tag{A.8}$$

where A is a positive constant, then we have, according to Dix [12], the following theorem;

**Theorem A1.** Assume that T(t) is a differentiable function satisfying eq(above), and that there exists a positive constant M, such that $|f'(t)| \leq M$ for all $t \geq 0$. Then $\lim_{t \to \infty} \dot{T}(t) = 0$.

From Lemma 1, we have $\dot{v} < 0$. Hence we establish that there is a bound for the derivative of $\dot{T}$. Therefore, the limit as t approaches infinity, as proven by Theorem A1, exists.

**Lemma 3 (Non-decreasing source energy).** If the system is SD, then the balance of energy is non-decreasing.

Proof. By definition it is straightforward to see that $\dot{T} < 0$. If $\dot{E} = 0$, then directly $\dot{E} - \dot{T} > 0$.